# Intrinsic shear transformations in metallic glasses

Zhen Zhang, Jun Ding[*], Evan Ma[*]

*Center for Alloy Innovation and Design, State Key Laboratory for Mechanical Behavior of Materials, Xi'an Jiaotong University, Xi'an, China.*

[*]Corresponding authors e-mails:   dingsn@xjtu.edu.cn (J.D); maen@xjtu.edu.cn (E.M)

**Abstract:** Plastic flow in amorphous solids is known to be carried by localized shear transformations (STs) which have been proposed to preferentially initiate from some "defect" units in the structure, akin to dislocations and point defects in crystalline solids. Despite the central role of STs in the mechanical deformation of metallic glasses (MGs), our knowledge of their characteristics has so far been limited to hypothetical models, based on computer simulations using unreleastically high cooling rates. Using combined molecular dynamics (MD) and Monte Carlo (MC) simulations, here we have succeeded in solidifying a metallic liquid at an effective cooling rate as slow as 500 K/s to approach that typical in experiments for producing bulk MGs. Exploiting this realistic MG model, we find that STs do not arise from signature structural defects that can be recognized *a priori*. Upon yielding, only about 2% of the total atoms participate in STs, each event involving as few as ~10 atoms. These findings rectify the unrealistically high content of "liquid-like regions" retained in MD-produced glass structures, which has rendered the MG model artificially ductile and under-predicted the sample-wide shear modulus by at least ~20% (with respect to that of experimental BMGs). Our finding sheds light on the scope of intrinsic structural inhomogeneity as well as the indeterministic aspect of the ST emergence under mechanical loading.

**Introduction**

Metallic glasses (MGs) are not only of practical interest due to their unusual mechanical properties [1-11], but also invaluable model systems for understanding the behavior of amorphous solids [6,12,13]. Generally speaking, a MG is expected to possess intrinsic heterogeneities in both its dynamics and internal local structure, which respond inhomogeneously to external stimuli such as a temperature rise and mechanical loading [14]. The basic event underlying the deformation of MGs involves the ST largely confined in an STZ [1-3]; the nature of STs has been extensively investigated over the past decades [2,15-23]. The generally accepted picture depicts STZ as a local group of around 100 atoms (or about 1.5 nm in diameter) collectively rearranging under externally applied stresses, and these STZs are closely related to the liquid-like regions retained in the MG structure [6,24]. This common belief, however, is based on phenomenological models (such as the one by Johnson and Samwer [16]) and atomistic simulations (e.g., Ref. [17]), and difficult to verify via direct experimental observations.

In particular, atomistic (e.g., molecular dynamics, MD) simulations carried out thus far in the MG field invariably suffer from the limitation that unrealistically high cooling rates had to be used (usually larger than $10^9$ K/s), leading to a long-standing concern that the computer-generated MGs would be very different from the experimental ones from liquid solidification [25-28]. For example, the shear modulus of any MD-simulated glass has always been far below that of the corresponding MG made in the laboratory. It is therefore obvious that the mechanisms derived from the MD simulations would not be entirely germane to real-world BMGs. To address this issue, efforts have been continuously made to bridge the gap between the quench rates in simulations versus those in experiments. Berthier and coworkers have done a series of work in this direction, by using the swap Monte Carlo (MC) algorithm [29] to produce highly stable configurations even down to the experimental glass transition temperature [30-32]. It was recently shown for two modified Lennard–Jones mixtures [33] that as the effective cooling rate is lowered, thermodynamic properties such as configurational entropy can be measured down to experimental $T_g$, and a qualitative change in the mechanical behavior of the glass (i.e., with increasing brittleness) can be observed [32].

However, a clear picture of the structural and mechanical inhomogeneities has remained elusive, especially when it comes to the characteristic features of the elementary plasticity carriers (i.e., STZs) and their behavior upon yielding of the MG.

Along this line of thought we begin with the following thought experiment: the structure of BMGs would be expected to become increasingly homogeneous (everywhere) with decreasing cooling rate of the parent liquid and/or extended relaxation/ageing below the glass transition temperature, eventually diminishing almost all "liquid-like regions" conducive to shear transformations (STs). As a result, the STZs responding to external stresses/loading would become far less populous than previously observed in the MD simulations, and very small in size towards the ~10 atoms known to be needed to trigger the STs [22,34], with little chance to cascade into the surrounding neighborhood. However, it has not been possible so far, experimentally or in simulations, to explore the scenario envisioned above. We can now reach this intrinsic limit, via modeling the MG using an approach combining molecular dynamics (MD) and Monte Carlo (MC) simulations. Through this method we cool the $Cu_{50}Zr_{50}$ liquid at an effective cooling rate as slow as 500 K/s to approach that typical in experiments for producing BMGs, aiming to produce new insight unreachable from the ultrafast quench used in all previous MD simulations.

The embedded atom method (EAM) potentials by Mendelev *et al*. [35] were used for describing the interatomic interactions. A typical melt-quench procedure under zero pressure was adopted for the MD-only simulations (cooling rate $\geqslant 10^9$ K/s). The hybrid MD/MC scheme adds a MC cycle at the end of every 10 MD steps, allowing for the swaps of local atoms during the MC steps, while the MD integration steps account for the relaxations. As such, the standard MD dynamics were supplemented with structural relaxation to an unprecedented level. Simulations were carried out in the variance constrained semi-grand canonical ensemble [36]. More details about the parameter setting, e.g., temperature profile and MD/MC integration, are given in the Supplemental Material (SM).

**Results and Discussion**

To begin with, we present in Fig. 1a the cooling curves, i.e., per-atom potential energy as a function of temperature. One recognizes that the coupling of MD and MC can significantly lower the energy state of the resulting glass relative to the MD-only method (note that the total simulation time increases from MD/MC-1 to MD/MC-4.). In addition, a glass transition (i.e., the deflection of the curve at intermediate temperatures) is seen in all cases, indicating that the MD/MC method indeed mimics the melt-quench procedure as it is usually adopted in conventional MD for producing model glasses, but with a much higher efficiency in accelerating the structural relaxation. Next, we estimate the effective cooling rates achievable using the hybrid MD/MC method through extrapolating the cooling rate-dependence of the potential energy of the MD-simulated $Cu_{50}Zr_{50}$ glass at 300 K. Figure 1b shows that a logarithmic fit describes very well the cooling rate-dependence of the MD-simulated $Cu_{50}Zr_{50}$ glass at 300 K (red-filled circles). Extrapolation of this trend to lower cooling rates indicates that the hybrid MD/MC method has yielded very slow effective cooling rate, down to $5\times10^2$ K/s for the small MG samples (containing 14,000 atoms) and $10^4$ K/s for the largest samples (containing 1,000,000 atoms). This was achieved at a moderate computational cost, demonstrating the high computational efficiency of the applied MD/MC method. It is worth noting that we detected no sign of crystallization in all the simulated samples, through a careful analysis of the local environment of each atom up to the second neighbor shell [37], affirming that all the computer MGs we have obtained are fully amorphous.

This dramatically reduced effective cooling rate, towards a level almost equivalent to that under laboratory experimental conditions for producing BMGs, yields glass models that are more realistic in terms of their macroscopic properties. Figure 1c shows that their total X-ray structure factor, $S(q)$, of simulated $Cu_{50}Zr_{50}$ MGs with different cooling rates. One notices that the agreement between the simulated and the experimental $S(q)$ [38] becomes progressively better, notably the first and the second peak, with decreasing cooling rates. Furthermore, Figure 1d shows that the shear modulus $G$ of the simulated MG continuously increases with lowering cooling rate. Previously, all the MD-simulated MGs are at least 20% smaller in $G$ compared with experimental data, due primarily to the unrealistically high cooling rates (see, e.g., Ref. [10]). Remarkably, this 20% deficiency of $G$ is gradually rectified in the MD/MC-simulated glasses as the effective cooling rate is lowered to the level comparable to that typical in experiments. Note that we are not shooting for a precise value of $G$, since on one hand the $G$ estimated from simulations could be interatomic potential-dependent, and on the other hand, the experimental value of *G* ranges from 24 to 31 GPa, depending on the geometry of the tested specimen and the technique used for the measurement [39,40].

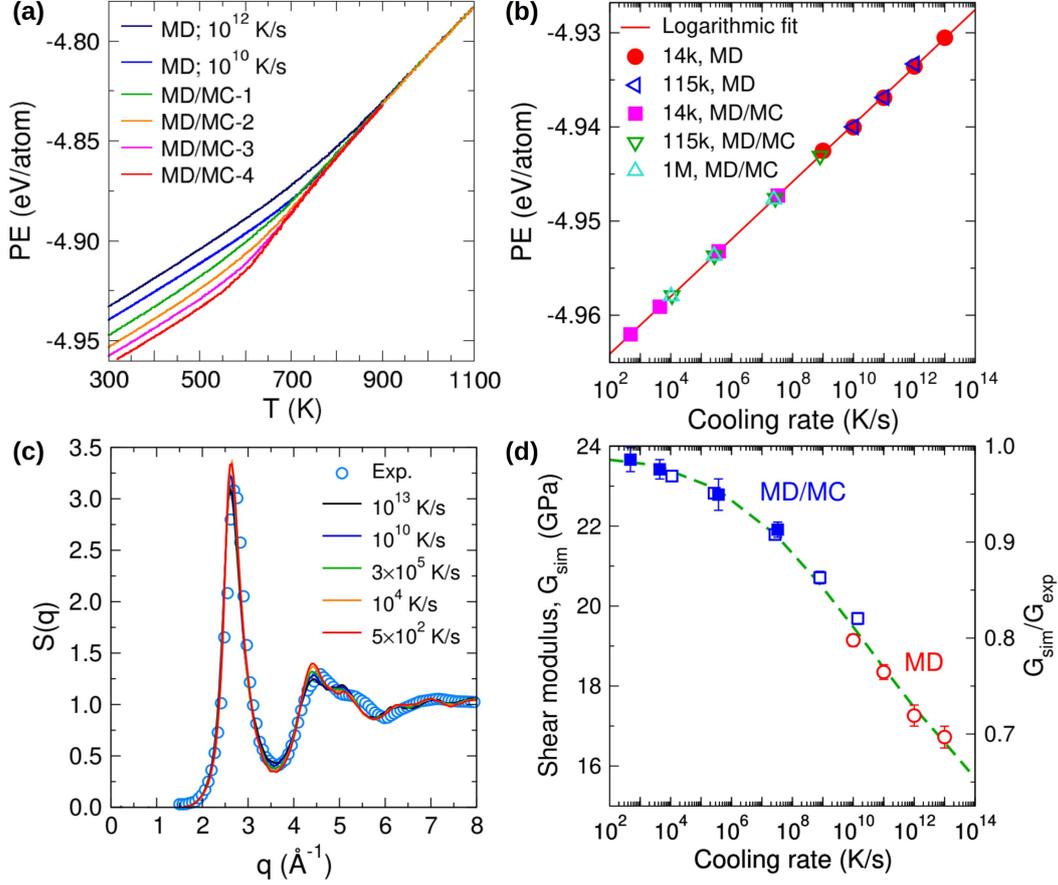

**Figure 1**. **$Cu_{50}Zr_{50}$ MG models prepared using various effective cooling rates.** (a) Variation in per-atom potential energy (PE) with temperature for the $Cu_{50}Zr_{50}$ MGs prepared using different effective cooling rates. (b) The per-atom potential energy estimated at 300 K vs. effective cooling rate. 14 k, 115 k, and 1 M represent, respectively, samples containing 14,000, 115,000, and 1,000,000 atoms. The solid line is a logarithmic fit to the MD data of the 14 k samples. Error bars are smaller than the symbol size. (c) Total weighted X-ray structure factor, *S*(*q*), at a variety of effective cooling rate. The experimental total X-ray structure factor for the $Cu_{50}Zr_{50}$ MG, from Ref. [38], is also included for comparison. (d) Simulated shear modulus $G_{sim}$ vs. effective cooling rate. Open and solid symbols are for the 14k and 115k samples, respectively. The dashed line is a guide to the eye. To calculate the ratio (see the right ordinate) between the simulated and experimental *G*, the latter is taken to be 24 GPa, measured from the stress-strain response of a $Cu_{50}Zr_{50}$ MG from Ref. [40] .

In SM Fig. S1, we provide additional results illustrating the glass transition temperature, fraction of Cu-centered full icosahedra, boson peak, uniaxial tensile strength, shear localization behavior, and radial distribution function. All of these properties exhibit a smooth cooling rate dependence over 10 decades, indicating that the MD/MC method can not only produce MG models that are highly consistent with the MD-simulated ones, but also enable the coverage down to processing conditions on par with those in lab experiments.

The influence of cooling rate is also obvious at the microscopic scale, in terms of the distribution of various atomic-level properties that can serve as indicators of the local structural environment and reflect the inhomogeneity of the latter. In Fig. 2, we show two representative local properties at 300 K, namely the vibrational displacement, $\lambda$ (top panels), which reflects the local configurational constraints, and the thermal activation energy, $E_{act}$ (bottom panels), as a metric for the ease of thermally-activated relaxation (simulation details are given in SM), for three MG samples prepared with an effective cooling rate of $10^{13}$ K/s (hyper-quenched, via MD), $10^9$ K/s (intermediate, via MD), and $5\times10^2$ K/s (slowly-cooled, via MD/MC).

Firstly, we note that these two local properties correlate quite well with each other, i.e., regions with larger $\lambda$ tend to have smaller $E_{act}$, in agreement with the finding of previous simulation studies [41]. Importantly, with decreasing cooling rate the distribution of $\lambda$ and $E_{act}$ become progressively less inhomogeneous (i.e., the distribution profile becomes narrower). $\lambda$ shifts toward smaller values while $E_{act}$ becomes larger, indicating that the glass structure becomes more rigid and less prone to structural relaxation. This finding is consistent with the annealing-induced homogenization of MG as revealed in experiments [42]. It is also interesting to note that the distribution profiles are clearly asymmetric, although the characteristic tail becomes less and less prominent with decreasing cooling rate, a result that implies a decreasing concentration of soft spots [47] in the MG. In SM Fig. S2, we also observe that the spatial autocorrelation between the two local properties decreases monotonically with decreasing cooling rate. This result agrees with the finding of a recent experimental study that efficient packing of particles in hard-sphere systems is associated with a *short* correlation length [43].

The homogenization of the local properties promoted via slow cooling is expected to have direct consequences on the stability of the glass [44], as well as the deformation behavior of the MGs. As the glass structure becomes less and less inhomogeneous with the slowing down of cooling the parent liquid, the majority of liquid-like regions (the blue regions in the snapshots, Fig. 2a and 2c) have been relaxed away. As a result, the left-over white-to-blue regions in the snapshots (Fig. 2) appear to have become scarce when compared with the MD-only cases, and very small in size, mostly less than 10 atoms.

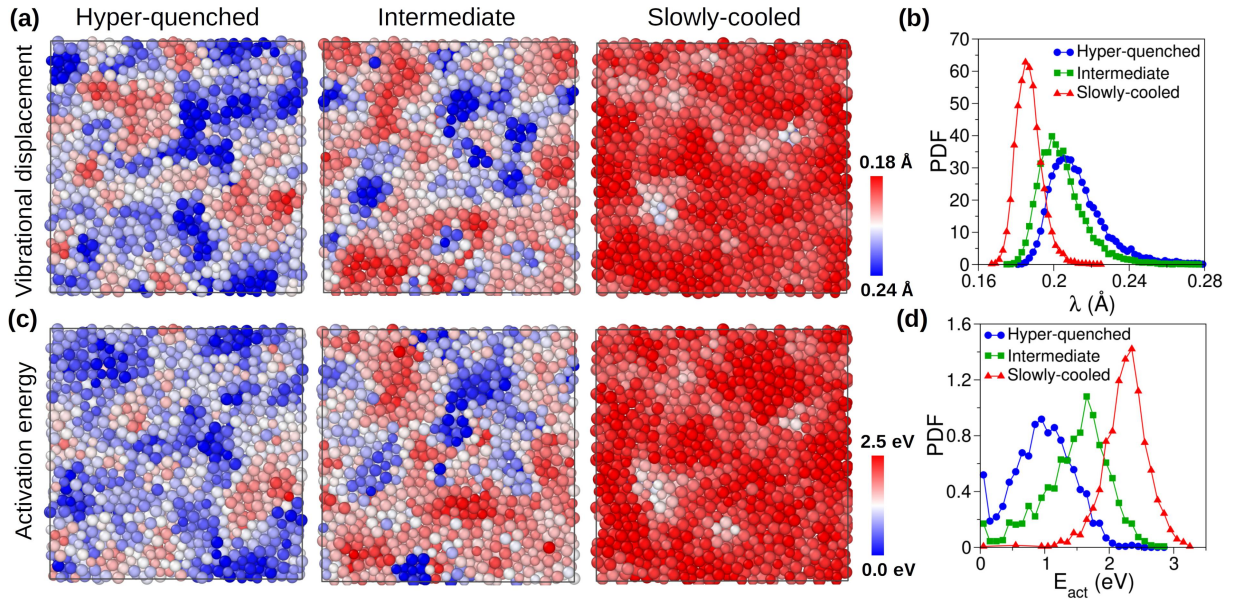

**Figure 2**. **Inhomogeneity in the MG samples.** Snapshots of the MG samples prepared via different effective cooling rates, mapping out the spatial distribution of (a) vibrational displacement, $\lambda$, and (c) thermal activation energy, $E_{act}$. The effective cooling rates are $10^{13}$ K/s, $10^9$ K/s, and $5 \times 10^2$ K/s, from left to right, respectively. (b) and (d) are the distribution of $\lambda$ and $E_{act}$, respectively. Both the actual width (i.e. the standard deviation) and the relative width (i.e., the standard deviation divided by the mean value) of the distribution profile decrease with decreasing cooling rate. The samples contain around 14,000 atoms, corresponding to cubes of side length ~6.5 nm at room temperature.

To explore the intrinsic deformation events in these well-relaxed glass, we have performed athermal quasi-static (AQS) shear simulations at a step size of $10^{-4}$ under conserved volume [45] (simulation details are given in SM). To begin with, we compare in Fig. 3a the stress-strain curves of three representative MGs (same as those in Fig. 2). One observes that, with decreasing cooling rate, the glass becomes increasingly stiffer, reflected by the increasingly steeper slope of the initial portion of the stress-strain curve. Importantly, with decreasing cooling rate the stress drops on the curve, which correspond to individual ST events, i.e., the elementary carriers of plasticity in MGs [3], become less in number and less pronounced in magnitude. An example of such a ST event involving 10 atoms is depicted in the inset of Fig. 3a.

Next, we use non-affine squared displacement, $D^2_{min}$ [3], i.e., the atoms with $D^2_{min}$ larger than a pre-defined threshold value $D_c$, as an indicator to identify the atoms that have participated in ST events. We have chosen $D_c = 1.2$ Å$^2$, based on the fact that the corresponding atoms are all in the exponential tail of the distribution of $D^2_{min}$ (see SM Fig. S3). We characterize the size of a STZ, $N_{STZ}$, as the number of newly appeared "shear transformed atoms" at each stress drop, see Fig. 3b. Figure 3c plots the distribution of the STZ sizes, which can be well described by a log-normal function (solid lines). Note that although this distribution depends on the threshold strains over which we count the STZs, Fig. S4, here we have used a sufficiently large strain of 0.08 to ensure a fair comparison between the various glasses. A significant reduction of both the number and the size of the STZs is apparent with decreasing cooling rate. For the hyper-quenched and the intermediate samples, the distribution of $N_{STZ}$ peaks at around $N_{STZ} = 10$ (not the mean value!), accompanied by a pronounced tail towards larger sizes. By contrast, for the slowly-cooled sample, not only is the number of the STZs significantly reduced, but the distribution of $N_{STZ}$ also becomes much narrower. The cumulative probability in the inset of Fig. 3c shows that 80% of the STZs in the slowly-cooled MD/MC sample have $N_{STZ} < 15$, whereas for the MD-hyper-quenched sample, a significant fraction of the STZs have $N_{STZ} > 100$. Figure 3d shows that the mean size of the STZs decreases from ~60 for the hyper-quenched sample to ~10 for the slowly-cooled sample and that the decrease of the mean STZ size with cooling rate can be well described by a power law. It is also worth noting that the spread of the STZ size distribution becomes increasingly narrower at lower cooling rate (and so does the relative width of the distribution profiles, see the inset), in accordance with the reduced degree of inhomogeneity in these samples (see. Fig. 2).

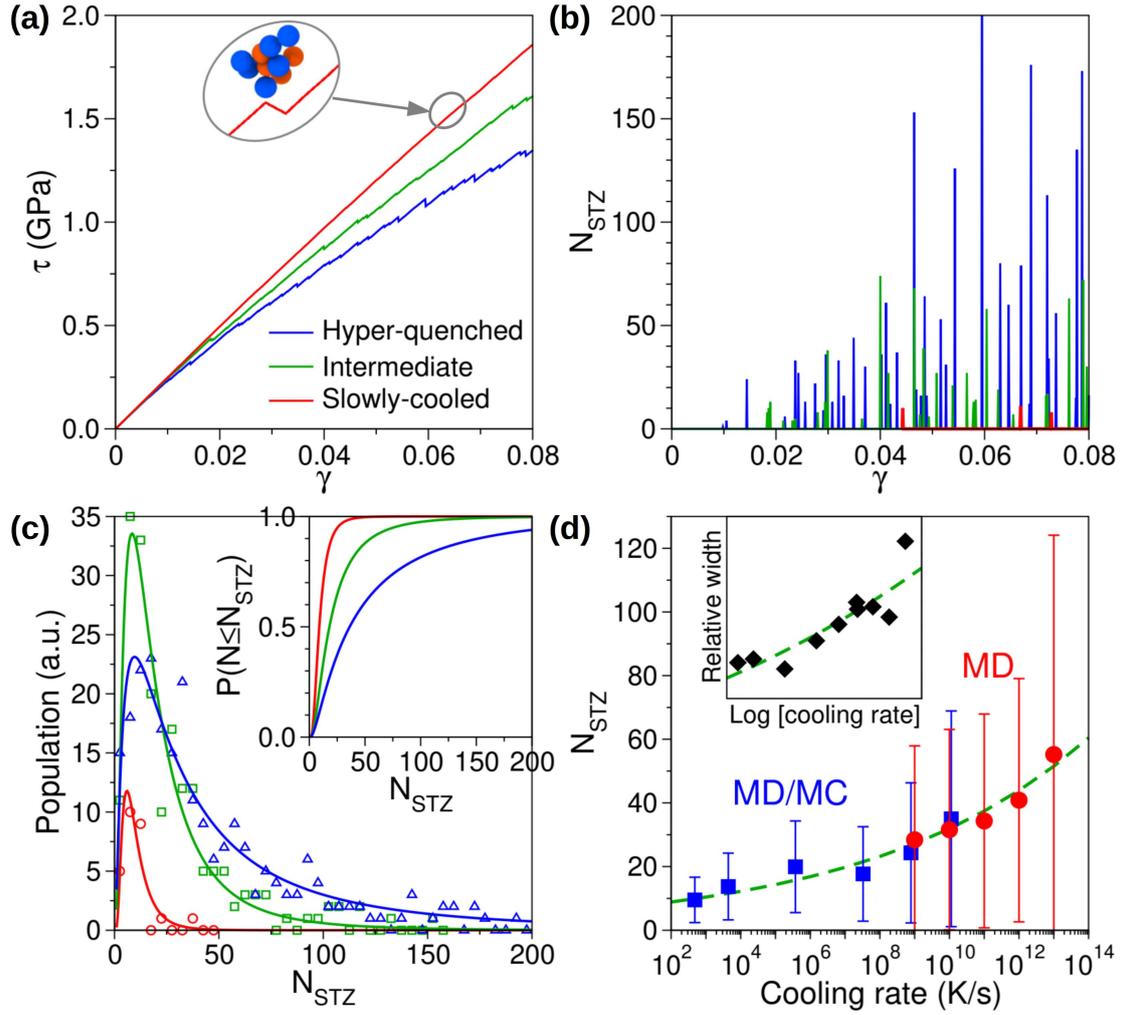

**Figure 3. Shear transformation zones in the MG samples.** (a) Shear stress ($\tau$) – shear strain ($\gamma$) curves for the athermal quasi-static shear of the MG samples in the $xy$ direction. The inset is the enlarged view of the stress drop at $\gamma_{xy} = 0.0669$ which is induced by the shear transformation of a cluster of 10 atoms (Cu and Zr are represented by balls in orange and blue, respectively). (b) The number of newly appeared "shear transformed atoms" at each stress drop, $N_{STZ}$, characterizing the size of a STZ. (c) Distribution of the STZ size. Solid lines are log-normal fits to the data points. The inset shows the cumulative probability $P(N \leq N_{STZ})$, i.e., the probability that the STZ size is smaller than or equal to the corresponding value on the x-axis. (d) Cooling rate dependence of the mean STZ size. Error bars denote standard deviation. The number of ST events are counted up to $\gamma = 0.08$. The inset shows the relative width of the distribution profiles in panel (c). Dashed lines are power-law fits to all the data points.

For the slowly-cooled sample, the STZ typically involves ~10 atoms. This is close to the number of atoms found to be needed for triggering thermally activated local rearrangement (also known as β relaxation) [22,34]. This finding suggests that in a well-relaxed glass structure the external stresses (e.g., via shear) would not be able to expand much in scope the trigger events for localized relaxation, in terms of the number of atoms and the size of the region involved. We also note that the STZ size identified here is only about one half of that estimated in previous studies (e.g., [21], through either a statistical analysis of the first pop-in stress during spherical nanoindentation, or the cooperative shearing model of Johnson and Samwer [16]).

Meanwhile, one would expect that these small clusters of atoms constituting the STZs would be spatially well separated from each other as each of them has a hard time invading into the surrounding neighborhood. In the following we show that this is indeed the case by analyzing the spatial distribution of the ST atoms. Figures 4a presents this distribution for the intermediate sample and the slowly-cooled sample, respectively, at the shear strain $\gamma = 0.08$. We observe that the ST atoms in the intermediate sample form percolated networks, whereas in the slowly-cooled sample these fertile sites only form small clusters and are isolated from one another. Also, the fraction of the ST atoms declines considerably from the intermediate MG to the slowly-cooled MG: Figure 4b shows quantitatively that the fraction of ST atoms decreased from ~25% in the hyper-quenched MG to only ~2% in the slowly-cooled MG.

Moreover, Fig. 4c shows for the ST atoms the spatial autocorrelation function, $C(r)$, which is defined as $C(r) = \frac{\langle P_{r_0} P_{r_0+r}\rangle - \langle P_{r_0}\rangle^2}{\langle P_{r_0}^2\rangle - \langle P_{r_0}\rangle^2}$. Here $P$ denotes $D^2_{\min}$, and $P_{r_0}$ and $P_{r_0+r}$ are, respectively, this property for an atom at a reference position $r_0$ and the value of the atom at a distance of $r$ from the reference point. One recognizes that the $C(r)$ decays in an exponential manner, which allows for fitting the data using the expression $y=A\times\exp(-x/\xi)$, where $A$ and $\xi$ are fitting parameters and the latter is the decay length. From the hyper-quenched sample to the slowly-cooled sample, the decay length $\xi$ is reduced by a factor of two, from 3.2 Å to 1.6 Å.

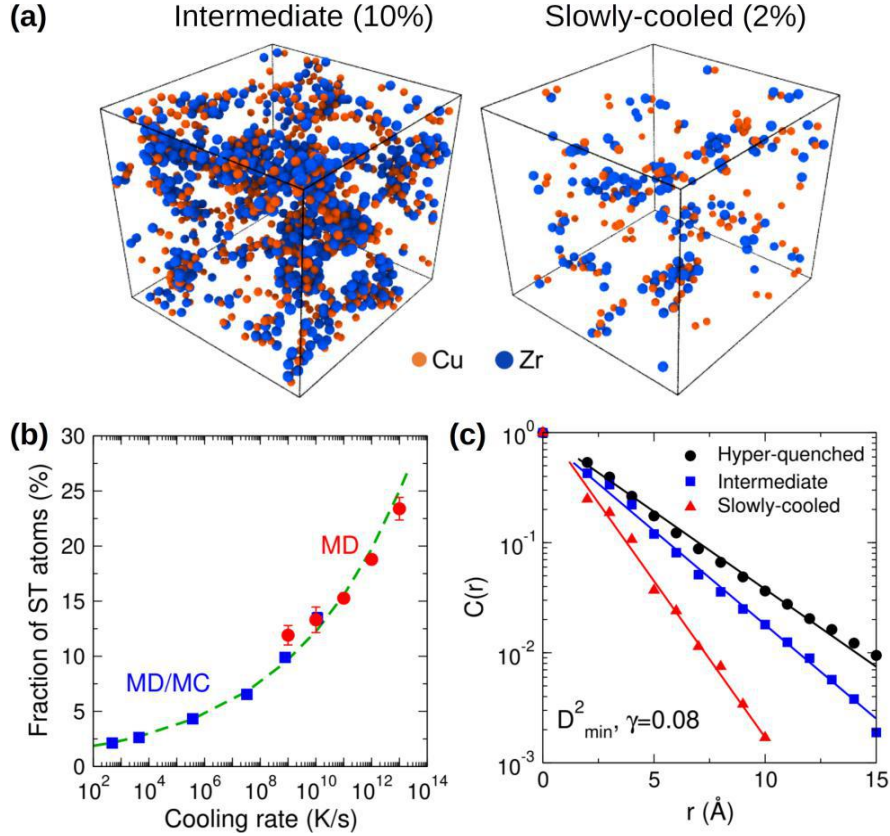

**Figure 4**. **Shear transformed atoms in the MG samples.** (a) Snapshots showing the spatial distribution of the ST atoms at the shear strain $\gamma = 0.08$. The number in the bracket indicates their fraction. (b) Cooling rate dependence of the fraction of the ST atoms. Dashed line is a power-law fit to all the data points. (c) Spatial autocorrelation function of $D^2_{min}$. Only the ST atoms are considered. Solid lines are exponential fits to the data points.

Taken together, these results indicate that all previous MD-simulated MG models have overestimated the fraction of fertile sites for STs and their percolation, inevitably making the glass models more ductile in deformation than real-world MGs. This conclusion is further supported by the deformation behavior of the MGs under uniaxial tension, which shows a gradual transition of the yielding from more ductile (spread-out of STs and shear bands, similar too previous MD simulations) to more brittle (higher propensity for strain localization and fewer shear bands) behavior with decreasing cooling rate (see SM Fig. S5), reminiscent of the effect of extended annealing of the initial sample [46].

Having produced MG models that are more close to experiments in terms of their macroscopic properties, a further important question to address is whether the pivotal sites for shear transformations (i.e., ST atoms) have distinct structure and properties such that they can be identified in the as-prepared static structure even before deformation? To shed light on this issue, we decompose the representative parameters of atomic properties that characterize the local structure of the slowly-cooled MG sample into contributions from the ST atoms and the rest.

Figures 5a shows that in terms of the distribution of vibrational displacement $\lambda$, the ST atoms exhibit notable difference from the rest sites, in that the distribution for the fertile sites peaks at a considerably larger $\lambda$ and has a pronounced tail to the right of the distribution. However, the overlap between the two distribution profiles are too strong to allow for a cutoff value that separates apart the two kinds of atomic sites. Figure 5b shows that the local activation energy $E_{act}$ of the fertile sites are on average smaller than that of the regular sites, but that there is also a strong overlap between the two distributions. Consistent with earlier MD simulation studies on even more heterogeneous MG structures, the sites more fertile for ST (white regions in Fig. 2) still tend to arise from the quasi-localized soft spots [47] (see plot for ST atoms in Fig. 5a). In other words, the tendency remains that upon loading the STs preferentially arise from high-$\lambda$ sites than the rest (Fig. 5a). These locations have higher flexibility volume [10] such that the atomic environments are more prone to rearrangements.

Regarding *bona fide* local structural quantities, Fig. 5c shows that the atomic volume for the ST-fertile sites are simply indistinguishable from that for the regular sites. In other words, there are no recognizable (e.g., nanometer sized) "loosely-packed regions" relative to (a matrix of) "densely-packed regions". Putting it another way, there are no identifiable defects arising from a noticeable deficit in atomic packing density, let alone a Swiss-cheese-like density bifurcation [7-9]. The atomic coordination number, on the other hand, shows a discernable difference between the two kind of sites, Fig. 5d, particularly for the Cu atoms: While the most probable coordination number of regular Cu is 12, the fertile Cu atoms are more likely to be found to have a smaller coordination number with the most probable being 11. These observations can be rationalized by their different chemical short-range order (SM Table S1). We find that the fertile Cu atoms are more likely to have Zr in their nearest neighbor shell than for the case of regular Cu; the larger size of Zr could thus compensate for the volume deficit of the fertile Cu due to smaller coordination number. In Fig. S6, we show that these observations of the insignificant differences between the atomic structure of the two kinds of atomic sites are independent of cooling rate by which the MG sample was prepared.

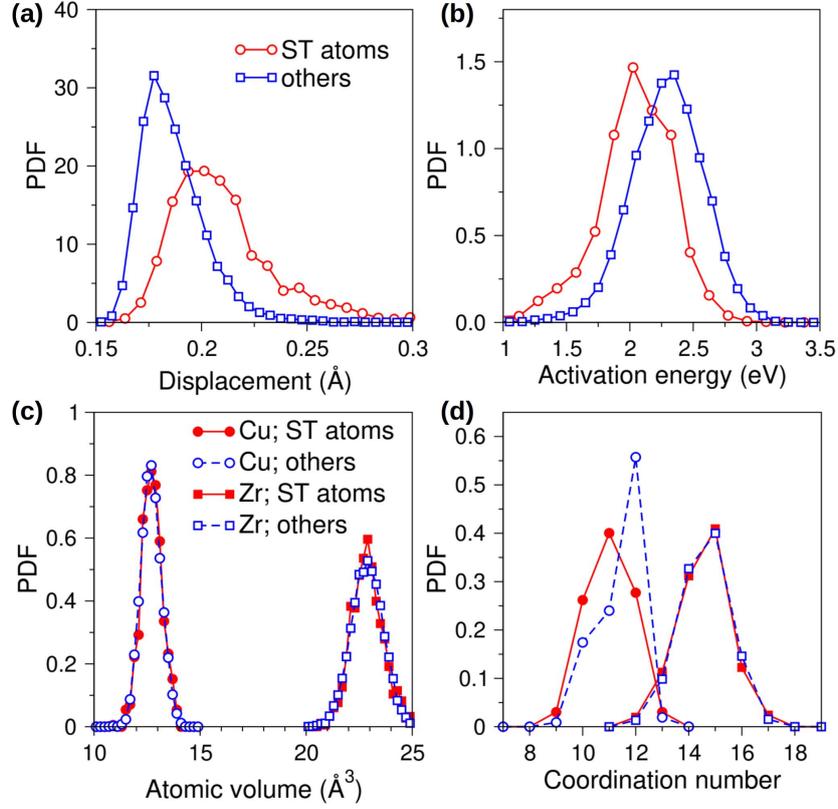

**Figure 5. Behavior of structural indicators in the MG samples.** Probability density function (PDF) of (a) the vibrational displacement, (b) local activation energy, (c) atomic (Voronoi) volume, and (d) coordination number. The number ratio between the shear transformed Cu and Zr is around one. These results are for the slowly-cooled MG (i.e., via the MD/MC method).

**Summary and concluding remarks**

To summarize, by implementing a hybrid MD/MC scheme, we have created a computer model of $Cu_{50}Zr_{50}$ MGs with an effective cooling rate as low as 500 K/s. With decreasing cooling rate of the parent liquid, the local structure and properties of the resulting glass become less and less inhomogeneous. Physics-informed local properties such as atomic vibrational amplitude and thermal activation energy, featuring a tail in their distribution profiles, are more sensitive than purely local structural parameters in revealing the inhomogeneity intrinsic to the MG. The cooling rate dependence of the two main characteristics of STZ, i.e., their mean size and concentration, can be well described by a power-law, i.e., a saturation of the shear transformation propensity is in sight. For the slowly-cooled MG which is very close to this saturated state, only 2% of the atoms are found to be fertile sites participating in the shear transformations at the shear stress of ~2 GPa, far less populous than previously believed. The STZs are very small in size, towards the ~10 atoms known to be needed to trigger the thermal activated deformation/relaxation [22]. This markedly smaller STZ is due to the increased constraints in the well-relaxed MG for the STZ to expand into its surrounding neighborhood. These findings remove some of the artifacts associated with previous super-fast MD quenching, and call into question some existing formulations, with regard to the STZ parameters used in currently prevailing models in the field.

Through detailed structural analysis, we have found that the atomic volumes of the sites participating in STs are indistinguishable from those of all the other atoms in the sample. Our MD/MC route clearly establishes the direction that intrinsically the MG structure would spontaneously eliminate the high-energy "loosely-packed regions". Hence there are no easily recognizable (e.g., from a few angstroms to nanometer in size) defects having markedly lower atomic packing density than the "densely-packed regions", let alone an extreme bifurcation into a Swiss cheese-like contrast purported in some papers [7-9]. We conclude that while the MG structure is non-uniform from location to location, the range/spread of inhomogeneity is not as wide as previously believed via MD-alone simulations. As a consequence, it is not possible to define atomic flow units that are responsible for plasticity, based on a *simple* parameter of the static structure inside an as-prepared MG. This is in accordance with an earlier view that STZs are not pre-existing defects that can be identified *a priori* but instead emerge dynamically in response to specific deformation conditions [21,22,48,49]. In order to establish a correlation between the local static glass structure and its rearrangement propensity, a major step-forward would have to rely on drastically boosting the amount of information in the local environment, to include multi-faceted atomic packing features as well as influence from their spatial anisotropy. To that end, instead of previously used pair distribution functions (i.e., an average or 1D projection), we could change the input into a huge "matrix" of atomic distribution functions in all directions over the 3D space. Recent advances in deep learning makes it possible to handle such a computationally intensive task [50]: a rotationally non-invariant local structure representation, combined with convolutional neural network, achieves a high-fidelity prediction of the loading-orientation-dependent plastic susceptibility from static structures in amorphous solids for stress-driven shear transformations.

Our MD/MC glass can be regarded as a benchmark computer model for well-relaxed MGs, i.e., a baseline BMG, into which additional heterogeneities can be introduced to appropriate degrees to compare with specific laboratory BMGs. For the latter, many extrinsic factors can be involved in solidification/processing that may heighten the heterogeneity in the MG. For example, casting BMGs in the lab would always involve temperature gradient, especially near the contact with mold surfaces, as well as gravity effects and non-uniformity associated with the fluid flow in the crucibles. Also, nanoscale fluctuations in chemical composition may be inevitable, especially since fast cooling is the norm when making glasses. Moreover, the potential sites/regions for atomic rearrangements may also be rendered more fertile around blemishes such as surface steps, impurity solutes, inclusions and casting defects, which may serve as heterogeneous nucleation sites for STs and for shear band initiation.

**Acknowledgements**

Z.Z., J.D. and E.M. acknowledge Xi'an Jiaotong University for hosting their work at the Center for Alloy Innovation and Design (CAID). Z.Z. and J.D. are partially supported by the National Key Research and Development Program of China under Grant No.2019YFA0209900; the National Natural Science Foundation of China under Grant No. 12075179.

# Supplementary Materials for:

# Intrinsic shear transformations in metallic glasses

In this Supplementary Material (SM) we provide additional details of the simulations, the methods for calculating various properties, and extended discussion/figures to support the conclusions in the main text.

## 1. Supplementary text

### 1.1 Simulation of the MGs

Our molecular dynamics (MD) simulations used the embedded atom method (EAM) potentials optimized recently by Mendelev *et al*. [35] which is suitable to simulate amorphous Cu–Zr alloys even at relatively low cooling rates. We have produced MG samples of $Cu_{50}Zr_{50}$ via the conventional MD or a hybrid MD/MC method [36].

For the pure MD scheme, a typical melt-quench procedure was adopted for obtaining the glasses with various system sizes. Periodic boundary conditions (PBCs) were applied in all three dimensions. A cubic box containing $N$ atoms was firstly melted at 1500 K for 0.5 ns to erase the memory of the initial positions of the atoms. Then the temperature was lowered to 1200 K and kept at this temperature until the sample (in the liquid state) reaches equilibrium, which was confirmed by examining the mean squared displacement (MSD) which showed that the liquid has entered the diffusive region and the mean MSD value is greater than 1000 Å$^2$ (corresponding to a simulation duration of ~1.0 ns). The equilibrated liquid sample was then cooled down from 1200 K to 300 K at various cooling rates (in the range from $10^9$ to $10^{13}$ K/s). For the slowly-quenched samples, we have used a two-step cooling protocol to save computation time: The cooling before 900 K was done with $\gamma = 10^{10}$ K/s and below this temperature the nominal cooling rate was applied. We note that this protocol does not affect the glass transition process as before 900 K the liquid is still in equilibrium and its properties are independent of the cooling rate used in this study. After quenching, the glass sample was further relaxed at 300 K for 2.0 ns, which is sufficiently long to ensure the convergence of the glass properties. System sizes range from $N$ = 14,000 to 1,000,000 were considered in our simulations. Five independent melt-quench runs were performed with various cooling rates for the samples consisting of 115,200 atoms. The error bar was estimated as the standard error of the mean of the five runs. The isothermal-isobaric (NPT) ensemble under zero pressure was used throughout the melt-quench process.

The hybrid MD/MC approach used a novel algorithm described in Ref. [36]. This hybrid scheme allows for local atom type swaps during the MC steps whereas the relaxations are accounted for by the MD integration steps. As such, the standard MD dynamics was interrupted at the reward of more efficient structural relaxation. Simulations were carried out in the variance constrained semi-grand canonical (VC-SGC) ensemble with the variance parameter $\kappa$ chosen to be 1000. One MC cycle containing N/4 trial moves was initiated every 10 MD steps. The MC temperature which determines the Metropolis acceptance criterion was set to be the same as the MD temperature. We note that the chosen parameters are not necessarily the optimal but instead for illustrating the usefulness of this hybrid scheme for producing highly relaxed MGs.

This hybrid scheme has been efficiently integrated in LAMMPS [51] and we have included an example input file for producing $Cu_{50}Zr_{50}$ samples with this method in the SM. Our simulations showed that this MD/MC scheme is as efficient as the pure MD scheme in terms of computational cost. However, the former can drive the system into equilibrium using a much smaller number of MD steps at a given temperature relative to the stand-alone MD method. Various nominal cooling rates were adopted depending on the simulated system sizes and the computational time required. In the cases where the melt-quench process was done by the hybrid MD/MC scheme, the as-obtained glass samples were further relaxed using the pure MD scheme at 300 K for 2.0 ns before evaluating the properties of the glasses.

**1.2 Mechanical deformation of the MGs**

Three types of mechanical deformation were considered in this study, namely volume conserved simple shear, athermal quasi-static shear (AQS) [45], and uniaxial tension. For the simple shear simulations, a strain rate of $5\times10^{-5}$ ps$^{-1}$ was applied for the *xy*, *xz*, and *yz* shearing up to a total shear strain of 0.02. The canonical (NVT) ensemble was used and 3D PBCs were applied for all simulations. The shear modulus *G* was calculated by a linear fitting of the stress-strain curve at small strains ($\gamma_{ij} \leqslant 0.01$, where $ij$ denotes the deforming direction). All simulations were performed at 300 K.

AQS simulations were carried out to probe the occurrence of the shear transformation (ST) events during the deformation of the MGs. The samples containing 14,400 atoms were used for the simulations. Simple $xy$ shear was performed at a step size of $10^{-4}$. After each step, the displaced configuration was energy minimized using the conjugate gradient algorithm before imposing further deformation to the sample. We used the non-affine squared displacement, also known as $D^2_{min}$ [3], as an indicator of local STZs, and the atoms with a $D^2_{min}$ value greater than 1.2 Å$^2$ were treated as "shear transformed atoms".

Uniaxial tension simulations were performed to investigate the formation of shear bands and thus the ductility of the MGs. The glass samples containing 14,400 atoms were replicated 5 and 10 times along the $x$ and $y$ directions, respectively, to produce large samples of dimensions 32 nm×64 nm×6.4 nm ($x$-$y$-$z$). These samples were firstly annealed at 550 K for 0.2 ns and then quenched to 300 K for another 0.1 ns of relaxation. This annealing-relaxation process was performed in the NPT ensemble under zero pressure, and with 3D PBCs applied. We note that this procedure is useful in minimizing the effect of the introduced boundaries from the replication of small samples. Next, the PBC along the $x$ direction was released and the sample was further relaxed at 300 K for 0.1 ns under the microcanonical (NVE) ensemble. Finally, the samples were subjected to uniaxial tension at the temperature of 300 K along the $y$ direction with a strain rate of $5\times10^7$ s$^{-1}$. We have also performed full melt-quench and tensile simulations for a sample of the same dimensions as the replicated ones described above, for the cooling rate of $3\times10^7$ K/s. No apparent difference in the stress-strain and shear banding behavior was observed. This justifies that the replication of small sample in our tensile simulations does not result in artifacts in the deformation behavior of the simulated glasses, but rather is an effective operation to reduce the computational cost of generating big glass samples.

All simulations described above were carried out using the Large-scale Atomic/Molecular Massively Parallel Simulator software (LAMMPS) [51]. Temperature and pressure were controlled using a Nosé–Hoover thermostat and barostat [52-54]. The MD time step was chosen to be 2.0 fs.

### 1.3 Calculation of atomic-level properties

We have evaluated two atomic-level quantities for quantifying the inhomogeneities in the simulated MGs, namely the vibrational displacement, $\lambda$, and local activation energy, $E_{act}$. The vibrational displacement reflects the local configurational constraints and was calculated according to the equation

$$\lambda_i = \sqrt{\langle |r_i(t) - r_{i,\text{equil}}|^2 \rangle_{\tau_0}},$$

where $r_i(t)$ is the position of atom $i$ at time $t$ and $r_{i,\text{equil}}$ is the equilibrium position of atom $i$. $\tau_0$ is the time interval over which we take the average. For including enough periods of oscillations we have chosen $\tau_0 = 100$ ps at 300 K. The vibrational displacement calculated on this time scale contains only the vibrational (i.e., without the diffusional) contribution.

To explore the local potential energy landscape (i.e. the potential energy minima and the saddle points), we have employed the ART nouveau method [22,34,55,56]. To study the local excitations of the system, initial perturbations in ART were introduced by applying random displacement on a small group of atoms (an atom and its nearest-neighbors). The magnitude of the displacement was fixed, while the direction was randomly chosen. When the curvature of the PEL was found to overcome the chosen threshold, the system was pushed towards the saddle point using the Lanczos algorithm [57]. The saddle point is considered to be found when the overall force of the total system is below 0.01 eV Å$^{-1}$. The corresponding activation energy is thus the difference between the saddle point energy and the initial state energy. For each group of atoms, we employed ~100 ART searches with different random perturbation directions. Since there were at least 10,000 such groups in each of our models, more than one million searches by ART were generated in total. After removing the failed searches and redundant saddle points, ~200,000 different activations, on an average, were identified for each of the samples.

### 1.4 Extended/supporting data

In Fig. 1 of the main text, we have shown that as the effective cooling rate via the MD/MC method is approaching the typical experimental cooling rate, the simulated glass structure becomes increasingly close to the experimental MGs. Here we provide further evidence to support this assessment. Figure S1 shows that all properties exhibit a smooth cooling rate dependence over 10 decades. Firstly, in Fig. S1a, one observes that the glass transition, as evidenced by the deflection of the *PE*(*T*) curve at intermediate temperatures, is clearly visible for all samples, and the estimated glass transition temperature depends logarithmically on cooling rate. Figure S1c demonstrates that the boson peak (BP) which has been known as an *unusual* vibrational property for glasses is also seen in the MD/MC-simulated glasses. The fact that the BP shifts towards higher frequency and lower intensity with decreasing cooling rate implies the reduction of atomic sites that are prone to rearrangement in the glasses prepared via the MD/MC approach, in accordance with the observations in Fig. 2, Fig. 3, and Fig. 4 of the main text. Figure S1f shows that the standard structural quantity, i.e., total radial distribution function, *g*(*r*), depends only mildly on cooling rate. Nevertheless, one notices that the agreement between the simulated and the experimental *g*(*r*) becomes progressively good with decreasing cooling rate.

Figure S4 shows that the distribution of STZ size for the slowly-cooled glass exhibits a stronger strain dependence relative to the hyper-quenched glass. This could be attributed to the fact that, in the case of the slowly-cooled glass, more and more STs are gradually activated with increasing strain, and when the localized STZs become close by to each other they begin to have an increased chance to each expand into the surrounding neighborhood, which results in the increase of the number of atoms involved in the identified STZs. We choose to fix the strain at 0.08 (corresponding to a stress value of ~1.8 GPa for the slowly-cooled glass) for a fair comparison between the shear deformation of the various glasses (see Fig. 4).

In Fig. 3 of the main text, we have shown that when subjected to AQS shear deformation, the MD/MC-simulated glass is stiffer and more brittle than the MD-simulated glasses. In Fig. S5 we show that this is also the case when the glass sample is under uniaxial tension. The stress-strain curves in Fig. S5a indicate that the glass becomes increasingly stiff, strong (maximum strength shown in Fig. S1d) and brittle with the decrease of cooling rate. This trend is manifested on the microscopic scale by the transformation of the shear banding behavior from multi-shear-banding (which leads to pronounced ductility) to single narrow shear-banding (which results in brittle fracture), see the snapshots in Fig. S5b. Note that we show in the snapshots only the atoms with atomic shear strain $\eta^{\text{Mises}}$ [58] larger than 0.3, in accordance with previous studies [59], for the visualization of the shear bands. This process is quantitatively captured by the drastic increase of the shear localization parameter $\psi$ [59], defined as $\psi = \sqrt{\frac{1}{N}\sum_{i=1}^{N}(\eta_i^{Mises} - \overline{\eta}_0^{Mises})}$ with $\overline{\eta}_0^{Mises} = \frac{1}{N}\sum_{i=1}^{N}\eta_i^{Mises}$. Figure S1e shows that $\psi$ increases from ~0.2 to 1.2 as the cooling rate decreases from $10^{12}$ K/s to $5 \times 10^2$ K/s.

## 2. Supplementary figures

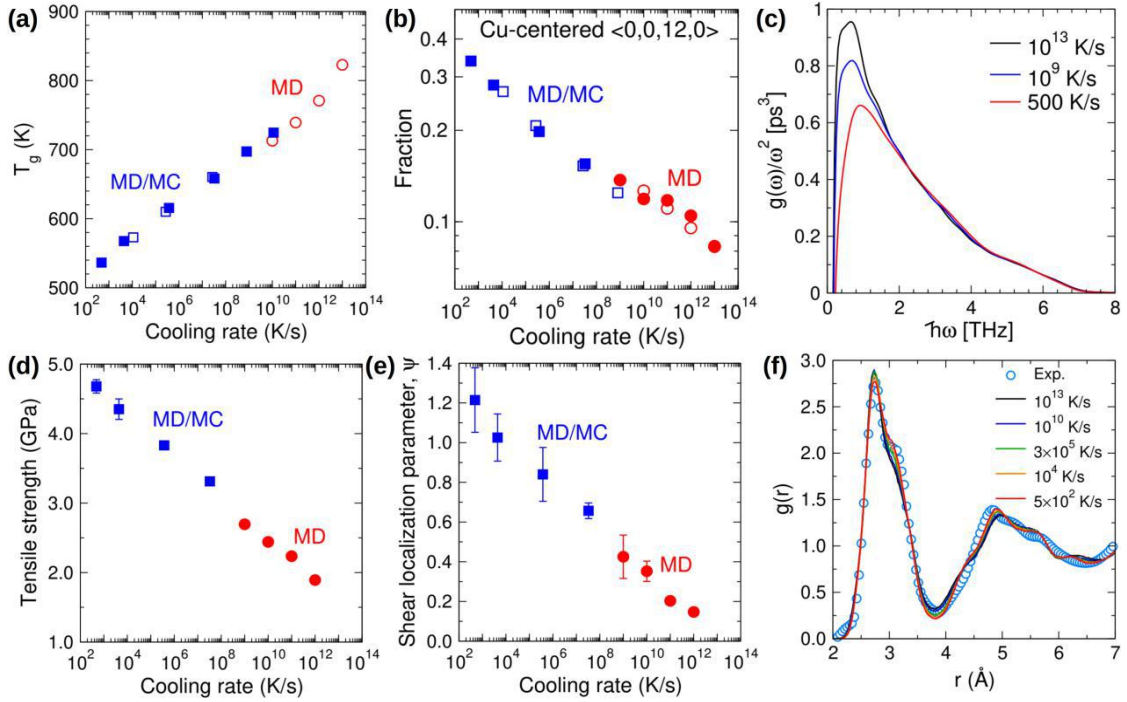

**Figure S1.** Cooling rate dependence of various glass properties. (a) Glass transition temperature $T_g$ estimated from intersection points in the PE($T$) curves. (b) Fraction of Cu-centered full icosahedra <0,0,12,0> (log-log plot). (c) Reduced vibrational density of state for the glasses showing that the boson peak shifts toward higher frequency and lower intensity with decreasing cooling rate. (d) Maximum strength of the glass under uniaxial tension. (e) Shear localization parameter $\Psi$. Open and filled symbols are for the samples containing 115,000 and 14,000 atoms, respectively. Error bars represent the standard error of the mean. (f) Total radial distribution function, $g(r)$. The experimental total X-ray pair correlation function for the $Cu_{50}Zr_{50}$ MG, obtained by the Fourier transformation of the total X-ray structure factor $S(q)$, from Ref. [38], is also included for comparison.

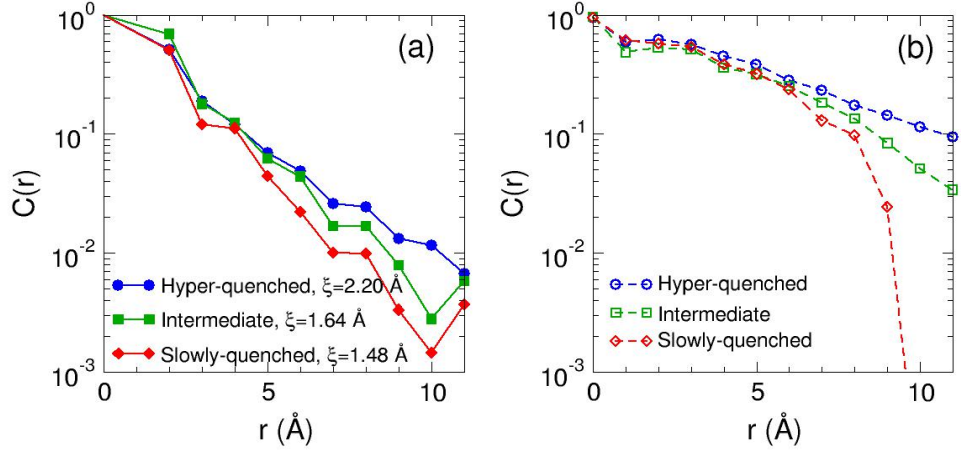

**Figure S2**. Spatial autocorrelation function $C(r)$ of (a) vibrational displacement, (b) local activation energy $E_{act}$. Note that the correlation functions for $E_{act}$ was measured in 2D (i.e., projection onto the $x$-$y$ plane) for a slice of 6 Å thickness perpendicular to the $z$-direction. The decay length $\xi$ for vibrational displacement was obtained by an exponential fit to the data using the expression $y=A*\exp(-r/\xi)$. $E_{act}$ does not exhibit a clear exponential decay and thus no decay length was estimated. One nevertheless observes the cooling rate effect, i.e., slowly cooled glasses show faster decay in their local properties.

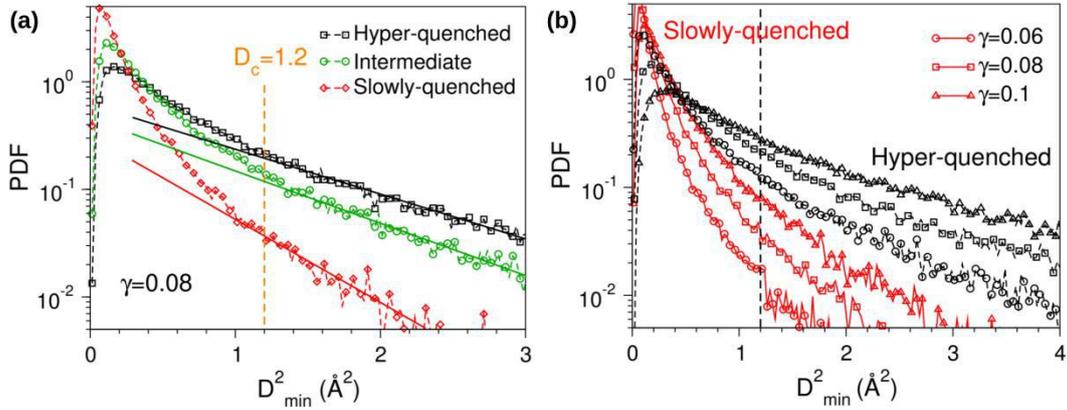

**Figure S3** (a) Probability density function (PDF) of non-affine squared displacement $D^2_{min}$ at $\gamma = 0.08$. The atoms in the exponential tail of the distribution, with $D^2_{min} > 1.2$ ($D_c$), are classified as shear transformed atoms. (b) This threshold value of $D^2_{min}$ depends only weakly on $\gamma$.

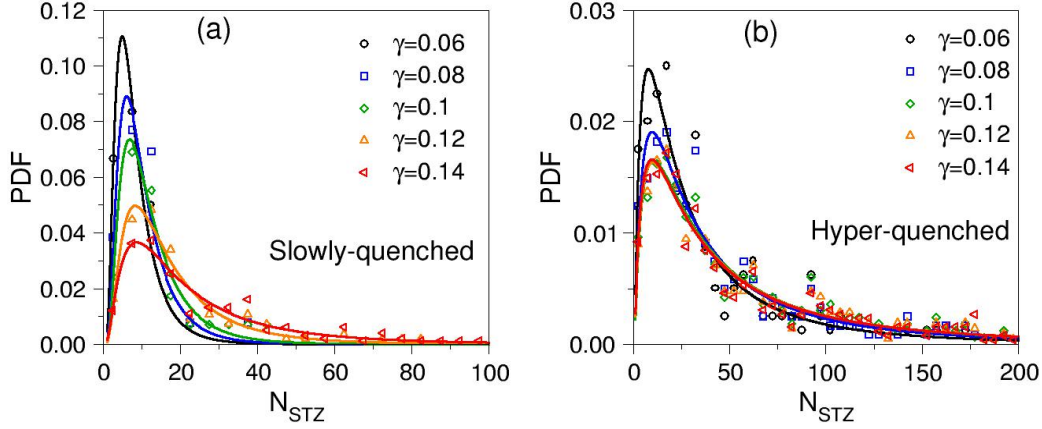

**Figure S4.** Probability density function (PDF) of the size of STZs, counted up to different shear strains experienced by the overall sample.

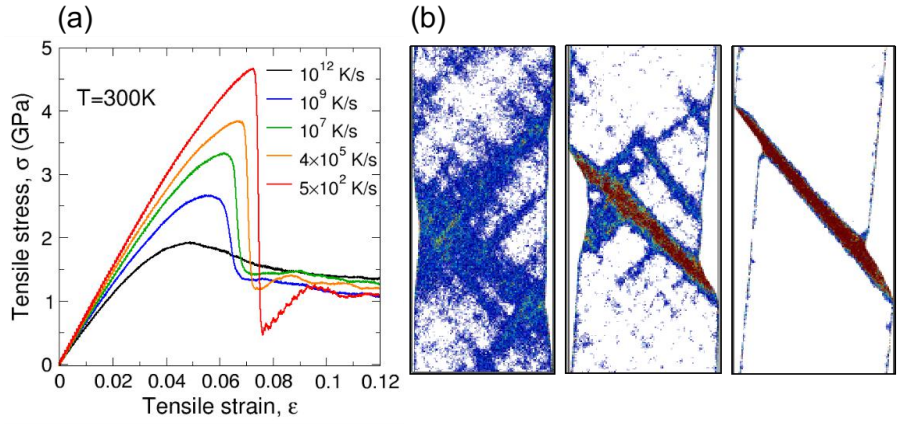

**Figure S5**. (a) Tensile stress-strain curve of the MGs prepared via different effective cooling rates at 300 K. (b) Snapshots of the MG samples at 12% of tensile strain. From left to right, the samples were prepared with a cooling rate of $10^{12}$ K/s, $10^9$ K/s, $5\times10^2$ K/s (via MD/MC). Atoms with von Mises local shear invariant $\eta^{\text{Mises}}$ less than 0.3 are deleted for better visibility of the shear bands. Color coding indicates atomic shear strain (ranges from 0.3~1.0). The calculated shear localization parameter $\Psi$ is shown in Fig. S1e.

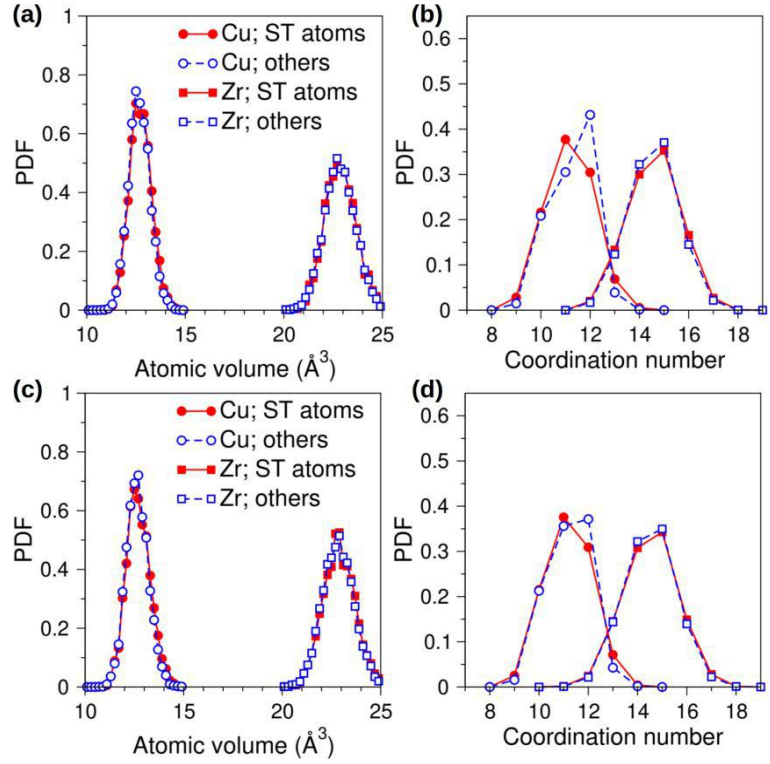

**Figure S6.** Atomic structure of the MD-simulated MG samples. (a-b) Probability density function (PDF) of atomic (Voronoi) volume and coordination number, respectively, of the intermediate-quenched sample. (c-d) The same quantities for the hyper-quenched sample.

| MD/MC: 5×10² K/s | | Cu | | Zr | | Cu | Zr |
|---|---|---|---|---|---|---|---|
| | | STZ | others | STZ | others | STZ/others | STZ/others |
| Atomic volume (Å³) | $\langle V \rangle$ | 12.7174 | 12.6915 | 22.8903 | 22.8791 | 1.0020 | 1.0005 |
| | $\sigma_V$ | 0.4804 | 0.4748 | 0.7575 | 0.7509 | | |
| total CN | $\langle Z \rangle$ | 10.9697 | 11.3817 | 14.6039 | 14.6352 | 0.9638 | 0.9979 |
| number of Cu | $\langle Z_{Cu} \rangle$ | 3.6450 | 4.4540 | 6.8676 | 6.9455 | 0.8184 | 0.9888 |
| fraction of Cu | $f_{Cu}$ | 0.3323 | 0.3913 | 0.4703 | 0.4746 | | |

**Table S1.** Statistics of the topological and chemical short-range order in the slowly-cooled $Cu_{50}Zr_{50}$ MG sample (via the MD/MC method). Atomic volume, total number of atoms in the first coordination shell, and the number of Cu in the first coordination shell are denoted by $V$, $Z$, and $Z_{Cu}$, respectively. $\langle X \rangle$ and $\sigma_X$ are respectively the mean value and the standard deviation of $X$. The number ratio between the shear transformed Cu and Zr is around one.